\documentclass{article}
\usepackage{gsirep}

\newcommand{\D}{{\rm d}}
\newcommand{\ca}{{\cal C}_A}
\newcommand{\cx}[1]{{\cal C}_{#1}}
\newcommand{\exva}[1]{\left\langle #1 \right\rangle} 

\title{Cluster Production in Ultrarelativistic Heavy Ion Collisions}
\author{R\"udiger Scheibl and Ulrich Heinz \\
         D-93040 Universit\"at Regensburg }
\date{}
\begin{document}
\maketitle
\noindent
\ldots may prove an interesting tool to extract physical quantities
of the fireball, once the formation process has been understood. 
We assume an emission function of the form \cite{emission} 
with gaussian source parametrization and local thermal equilibrium
distributions for nucleons in the hot, central reaction region. 
Using a Wigner function coalescence mechanism \cite{ich}, we obtain a 
formula for the invariant momentum spectra of clusters of $A$ nucleons
\cite{ich2}:
\begin{eqnarray}
\lefteqn{
E \frac{\D^3 N}{\D P^3}  = M\, \frac{2J+1}{(2\pi)^3}\, 
\exp\left(\frac{\mu_A-M}{T}\right)\, \exva{\ca} V_{\rm eff}^{(A)}  
} \nonumber \\ & & \times 
\exp\left(- \frac{M_t-M}{T_*} - \frac{A\, Y^2}{2 (\Delta\eta)^2}  \right) 
\enspace , \quad\quad\quad\quad  \label{f1}
\end{eqnarray}
where the first factors are the cluster mass $M=A\,m$ 
($m$ is the nucleon mass), the cluster's spin degeneracy over phasespace, 
and the fugacity with the chemical potential $\mu_A$ of the cluster. 
The coalescence correction $\exva{\ca}$ contains the quantum mechanis 
of coalescence and basically relates the cluster wavefunction to system 
inhomogeneities due to flow (for nucleons, $\exva{\cx{1}}\equiv1.0$). 
Flow also causes a strong correlation between momentum and spatial
coordinates of a particle, which makes the effective volume $V_{\rm eff}$
of the source of particles with certain momentum 
smaller than the overall fireball volume. In our model, $V_{\rm eff}$ 
is only slightly dependent on transverse momentum and nearly independent of
longitudinal rapidity $Y$, but scales with the cluster mass. Further,
it is related to the HBT-parameters $R_\parallel$ and $R_\perp$ in the 
YKP-parametrization \cite{emission}: 
\begin{eqnarray}
V_{\rm eff}^{(A)} & = & A^{-\frac{3}{2}}\, V_{\rm eff}^{(1)}  
\label{f2} \enspace ,\\
V_{\rm eff}^{(1)} & \sim & \frac{m_t}{m}\, R_\parallel(m_t)\, R_\perp^2(m_t)  
\label{f3} \enspace .
\end{eqnarray}
Lastly, $\Delta\eta$ is the width of the spacetime rapidity distribution 
and $T_*$ the effective temperature of the transverse momentum distribution. 
$T_*$ is calculated from the freeze-out temperature $T$ and the 
transverse flow rapidity $\eta_f$:
\begin{equation}
T_* = T + (M/A)\,\eta_f^2  \enspace .
\end{equation}  
Note that $T_*$ is independent of cluster mass: through $M/A=m$ only the
nucleon mass enters the formula. This is in contradiction with experiments,
where deuterons show much higher effective temperatures than protons 
(at least in the mid-rapidity region of central collisions, z.B. \cite{NA44}).
This can be cured by a box profile for the transverse fireball geometry
instead of a gaussian, and leads to a non-analytic formula instead
of (\ref{f1}) and to a different scaling in (\ref{f2}), e.g. 
$A^{-\frac{1}{2}}$. A more substantial study is on its way \cite{ich2}. 
From (\ref{f1}) follows a momentum-independent invariant coalescence 
factor $B_A$: 
\begin{equation}
B_A  = \exva{\ca}\ \frac{2J_A+1}{\sqrt{A}\, 2^A}\ 
\left( \frac{(2\pi)^3}{m}\ \frac{1}{V_{\rm eff}} \right)^{A-1}
\enspace .
\end{equation}
A box profile for the transverse fireball geometry will
result in a rise of $B_A$ with larger transverse momentum.
With $V_{\rm eff}$ approximately known from HBT analysis for central 
Pb+Pb collisions at the SPS and 
$\exva{\cx{2}}\approx\exva{\cx{3}}\approx0.7$ \cite{ich2},
we obtain for these reactions
$B_2 \approx 8\cdot 10^{-4}\,{\rm GeV^2} $ and 
$B_3 \approx 4\cdot 10^{-7}\,{\rm GeV^2}$ in reasonable agreement
with experiments \cite{NA44,NA52}. We further consider the ratio $S_{AA'}$
of the invariant momentum spectra (\ref{f1}) of different cluster species
at zero particle momentum 
$\vec{P}_A \!=\! \vec{P}_{A'}\!=\!\vec{0}$ 
(in the rest frame of the fireball). These particles are likely to come from 
the central collision zone.
From measured anti-particle/particle ratios $S_{\bar{A}A}$ 
the fugacity $\mu_A/T$ can be determined. 
Since $\exva{\ca}$ is only weakly dependent on flow and 
temperature, we can use an approximate knowledge of it to extract the 
freeze-out temperature $T$ of the fireball from the ratio of different 
cluster spectra:
\begin{equation}
T = \frac{(A'-A)\, m} 
         {\ln(S_{AA'}) - \ln\frac{A\,(2J_A+1)}{A'\,(2J_{A'}+1)} - \ln(\kappa) 
         + \frac{\mu_{A'} -\mu_{A}}{T} }
\enspace .
\end{equation}
The variable $\kappa$ contains the ratio of the $A$-depending effective
volumes and the coalescence corrections:
\begin{equation}
\kappa = \left(\frac{A'}{A}\right)^{3/2}\,\frac{\exva{\ca}}{\exva{\cx{A'}}}
\enspace .
\end{equation}
In a purely thermal model for cluster emission, $\kappa=1$. The later was
used by NA52 to extract a freeze-out temperature from their data on
minimum bias Pb+Pb collisions at 158\,GeV per nucleon and resulted in
substantially different temperatures from different particle ratios 
\cite{NA52}.  This contradiction no longer exists in our analysis: 
\begin{displaymath}
\begin{array}{|c|c|c|}
\hline
S_{AA'} & \multicolumn{2}{c|}{T\ {\rm [MeV]}  }            \\ \hline
 & \mbox{thermal model \cite{NA52}}  & \mbox{our model}  \\ \cline{2-3} 
{\rm p/d}              & 120.7 \pm 1.8 &   \\
{\rm \bar{p}/\bar{d}}  & 117.1 \pm 3.1 & \raisebox{7pt}[-10pt]{144} 
\\ \cline{3-3}
{\rm d/t   }           & 135.8 \pm 9.3 &   \\
{\rm \bar{d}/t}        & 139.5 \pm 4.9 & \raisebox{7pt}[-10pt]{146} 
\\ \hline
\end{array} 
\end{displaymath}
We used slightly different chemical potentials for neutrons and
protons (known from $\rm \bar{p}/p$ and $\rm \bar{d}/d$), which 
make the temperatures from ${\rm p/d}$ and ${\rm \bar{p}/\bar{d}}$
(${\rm p/t}$ and ${\rm \bar{d}/t}$, respectively) 
naturally coincide. This can be justified by the initial neutron excesss in 
the cold Pb ion, a remainder of which may still be found in the
central collision region at freeze-out.

\end{document}